\documentclass[runningheads]{llncs}
\usepackage{float}
\usepackage{hyperref}
\usepackage[T1]{fontenc}
\usepackage{amssymb}
\usepackage{graphicx}
\usepackage{color}

\begin{document}
\title{Reproducing Bayesian Posterior Distributions for Exoplanet Atmospheric Parameter Retrievals with a Machine Learning Surrogate Model
}
\titlerunning{Machine Learning for Exoplanet Parameter Retrievals}
%
\author{Eyup B.~Unlu\inst{1}\orcidID{0000-0002-6683-6463} \and
Roy T.~Forestano\inst{1}\orcidID{0000-0002-0355-2076} \and
Konstantin T.~Matchev\inst{1}\orcidID{0000-0003-4182-9096} \and
Katia Matcheva\inst{1}\orcidID{0000-0003-3074-998X} 
}
\authorrunning{E.~Unlu et al.}
%
\institute{Institute for Fundamental Theory, Physics Department, University of Florida, Gainesville FL 32653, USA}
\maketitle              
\begin{abstract}
We describe a machine-learning-based surrogate model for reproducing the Bayesian posterior distributions for exoplanet atmospheric parameters derived from transmission spectra of transiting planets with typical retrieval software such as TauRex. The model is trained on ground truth distributions for seven parameters: the planet radius, the atmospheric temperature, and the mixing ratios for five common absorbers: $H_2O$, $CH_4$, $NH_3$, $CO$ and $CO_2$. The model performance is enhanced by domain-inspired preprocessing of the features and the use of semi-supervised learning in order to leverage the large amount of unlabelled training data available. The model was among the winning solutions in the 2023 Ariel Machine Learning Data Challenge.

\end{abstract}
\section{Introduction}

The discovery of planets orbiting stars beyond our own has drastically reshaped our perception of the cosmos and our role in it. With the advancements in observational techniques and the increasing number of space-based missions, an ever-growing catalog of exoplanets has emerged, revealing an astounding range of varying planetary habitats. Invaluable insights about the structure and composition of an exoplanet atmosphere come from transit spectroscopy, whereby a planet passes in front of its host star and the associated relative reduction in the original stellar flux $\Phi_0$,
\begin{equation}
M_\lambda \equiv \frac{\Phi_0 - \Phi_i}{\Phi_0},
\end{equation}
is measured in several different wavelengths $\lambda$ (here $\Phi_i$ is the minimum stellar flux during transit). 

However, extracting significant and dependable information from these observed transit spectra necessitates a comprehensive and thorough understanding of the underlying theory, observational limitations, numerical simulations, and statistical inference involved in the retrieval process.

In order to model the observed transit spectra, one employs complex forward radiative transfer models that account for the atmospheric structure, composition and planet parameters. The inverse problem of exoplanet parameter determination is typically tackled by Bayesian retrieval models, which, given an observed spectrum, provide the posterior distributions of the planetary atmospheric parameters. Bayesian retrievals are considered as the gold standard in the field, but are notoriously slow, since sampling-based retrievals typically require between $10^5-10^8$ calls to the forward model before reaching convergence \cite{Changeat_2022}. The arrival of the next generation of telescopes with high-resolution spectroscopic capabilities for exoplanet characterisation is projected to result in a substantial increase in the availability of high-resolution (i.e., high-dimensional from machine learning point of view) spectroscopic data for thousands of planets. The use of traditional data analysis and retrieval tools will be prohibitively slow in the face of the sheer volume of data and number of planets to be analyzed. This necessities the development of a new generation of fast, efficient, and accurate retrieval methods.

\section{Objectives}

Since 2019, the Ariel Machine Learning Data Challenge has been an annual hackathon event that seeks innovative machine learning (ML) solutions to pressing issues faced by the exoplanet community and the Ariel Space Telescope in particular \cite{Nikolaou_2020,Changeat_2022,Yip_competition,Ariel2023}. Similar to the James Webb Space Telescope, the Ariel mission will perform infrared spectroscopic observations of transiting exoplanets at high resolution. The specific task for the 2023 installment of the Ariel Machine Learning Data Challenge was to develop a surrogate ML model, which, given a simulated observation, can predict the posterior distributions for seven planet parameters: the planet radius $R_p$, the atmospheric temperature $T$ and five molecular abundances ($H_2O$, $CH_4$, $CO_2$, $CO$, $NH_3$).  The ground truth was the corresponding posterior distribution derived with the {\sc TauRex} suite \cite{Al-Refaie_2021} using a MultiNest algorithm \cite{feroz2008,feroz_2019} (for further details, see 
\cite{Changeat_2022,Yip_competition}).

\begin{figure}[t]
\begin{center}
\includegraphics[height=0.47\textwidth]{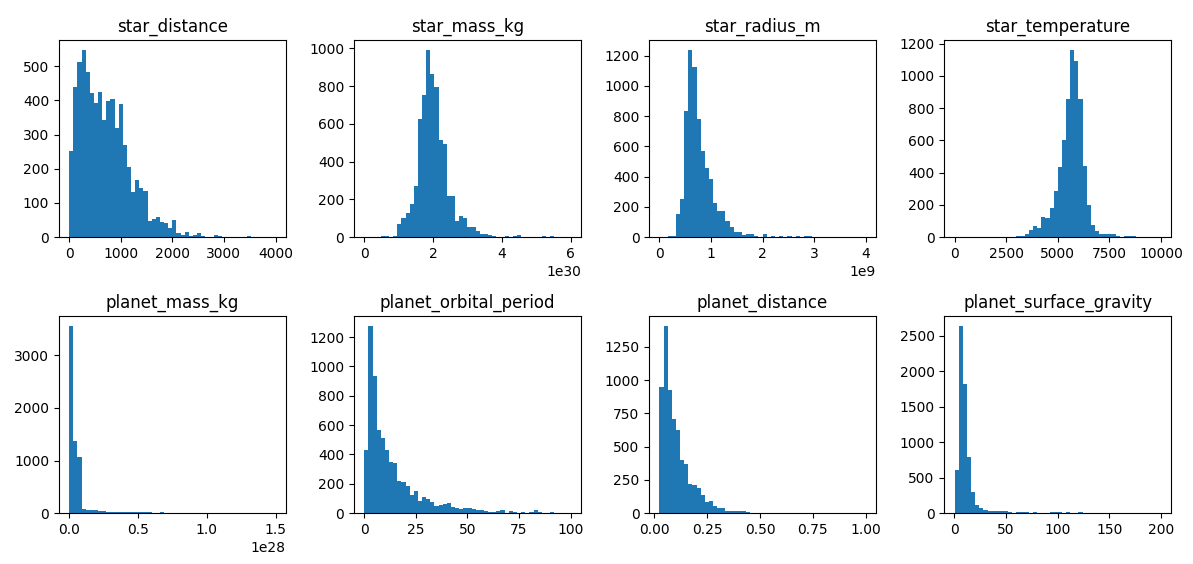}\\
\includegraphics[height=0.47\textwidth]{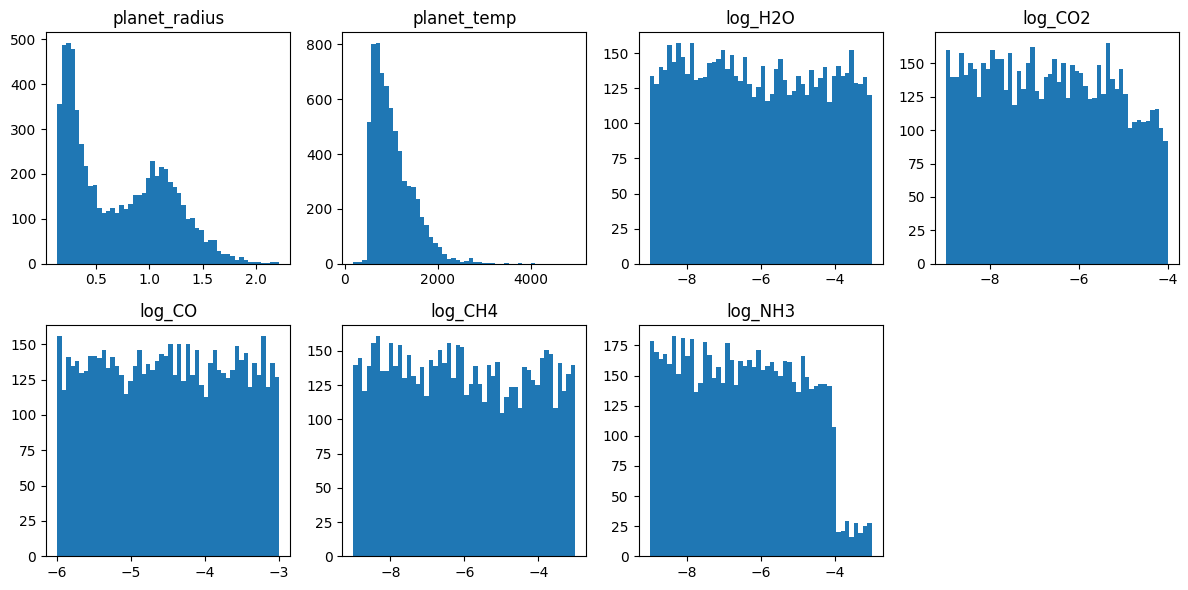}
\end{center}
\caption{Distribution of the auxiliary parameters (top eight panels) and the forward model parameters (bottom seven panels) in the training database. 
} 
\label{fig:database}
\end{figure}

The training dataset provided by the challenge organizers contained a total of 41423 planets. For each planet, the observed transit depth $M_\lambda$ was provided at 52 different wavelengths $\lambda$ between $0.55\mu m$ and $7.28\mu m$, along with the associated uncertainty $\epsilon_\lambda$ for each spectral bin $\lambda$. Additionally, the training dataset contained information about a number of auxiliary parameters describing the star-planet system, e.g., the star distance, the star mass, the star radius, the star temperature, the planet mass, the planet orbital period, the planet distance and the planet surface gravity. The distributions of those parameters over the entire dataset are shown in the top eight panels in Figure~\ref{fig:database}. The training database also contained the values of the forward model parameters used to produce the synthetic transit spectrum for each planet. Those parameters were the planet radius $R_p$, the atmospheric temperature $T$, and the mixing ratios $X_i$ for five common absorbers, where $i\in \{H_2O, CH_4, NH_3, CO, CO_2\}$. Their distributions are shown in the bottom seven panels in Figure~\ref{fig:database}, where $R_p$ is given in units of the Jupiter radius $R_J$, $T$ is in degrees Kelvin ($K$), and the mixing ratios are on a log base 10 scale. The forward model parameters ranges are $0.075 R_J$ to $2.43 R_{J}$ for the planet radius $R_p$, $101K$ to $4965K$ for the atmospheric temperature $T$, from -9 to -3 for $\log X_{H_2O}$, $\log X_{CH_4}$ and $\log  X_{NH_3}$, from -6 to -3 for $\log X_{CO}$ and from -9 to -4 for $\log X_{CO_2}$. Note that while the forward model parameters were provided in the training dataset, and thus could be used during training, they were not available in the test data, and therefore should not be among the inputs of the trained surrogate ML model.

For 6766 of the planets in the dataset, posterior distributions derived with {\sc TauRex} were provided in the form of weighted samples. From now on we shall refer to those planets as labelled. The remaining 34657 planets in the dataset did not have such posterior distributions and are therefore unlabelled. 

\section{Methodology}

We develop three separate custom ML models to predict the chemical composition $X_i$, the planet radius $R_p$, and the atmospheric temperature $T$, respectively.

\subsection{Feature Engineering}

For all models, we expand the initial set of auxiliary parameters $\{Aux\}$ to a larger set $\{Aux+\}$ by adding i) an approximate estimate $\hat R_p$ of the planet radius and ii) four dimensionless features motivated by domain knowledge \cite{2022ApJ...930...33M,2022PSJ.....3..205M}
\begin{equation}
\{Aux+\} = \left\{Aux, \hat R_p,\frac{\hat R_p}{R_s},\frac{D}{H},\frac{R_s}{H},\max (M'_\lambda)\right\}.
\label{eq:aux}
\end{equation}
The minimum transit depth, $M_{min}\equiv \min_\lambda(M_\lambda)$, provides the most stringent upper bound on the planet radius, therefore we take $\hat R_p = R_s\sqrt{M_{min} }$ as a useful proxy for $R_p$. Another quantity of interest entering (\ref{eq:aux}) is the scale height of the atmosphere $H=\frac{k_BT_e}{mg}$, which depends on the Boltzmann constant $k_B$, the mean molecular mass $m$ (approximated as 2.29 amu for a typical gas giant), the specific gravity $g$, and the planet equilibrium temperature $T_e$. The latter can be estimated from the available auxiliary parameters as $T_e= T_s\sqrt{\frac{R_s}{2D}}$, where $T_s$ is the star temperature, $R_s$ is the star radius, and $D$ is the distance from the planet to the star.

Subsequently, the auxiliary information (\ref{eq:aux}) was standardized by subtracting the mean and dividing by the standard deviation over the training set for each individual parameter, respectively. 

In order to focus on the absorption of the atmosphere alone, we modify the transit spectrum $M_\lambda$ by subtracting the opaque disk of the planet as \cite{2022PSJ.....3..205M,2022ApJ...939...95M}
\begin{equation}
M_{\lambda}'=M_{\lambda} - \frac{\hat R_p^2}{R_s^2},
\end{equation}
and rescaling as 
\begin{equation}
M_{\lambda}'' = \frac{M'_{\lambda}}{\max_\lambda (M_{\lambda}')}.
\label{eq:rescaling}
\end{equation}
The noise data was rescaled in a similar manner:
\begin{equation}
\epsilon_{\lambda}' = \frac{\epsilon_{\lambda}}{\max_\lambda (M_{\lambda}')}.
\label{eq:rescaling_noise}
\end{equation}
Due to the rescalings in eqs.~(\ref{eq:rescaling}) and (\ref{eq:rescaling_noise}), we include $\max_\lambda (M_{\lambda}')$ in the extended list of parameters in (\ref{eq:aux}). 

In addition to (\ref{eq:aux}), (\ref{eq:rescaling}) and (\ref{eq:rescaling_noise}), the models for the temperature and the planet radius also used the unmodified spectra $M_\lambda$ among their inputs. In summary, the inputs for the radius and the temperature ML models were $(M_{\lambda},M_{\lambda}'',\epsilon_{\lambda}',Aux+)$, while the inputs for the chemical ML model were simply $(M_{\lambda}'',\epsilon_{\lambda}',Aux+)$.

\subsection{Posterior Distribution Parametrization} 

\begin{figure}[t]
\centering
\includegraphics[width=0.99\textwidth]{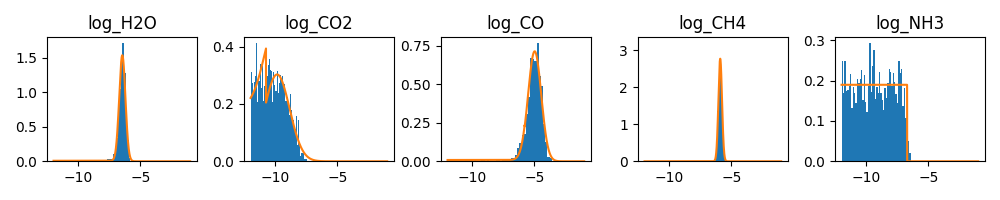}     
\caption{Comparison of the given ground truth distributions (blue histograms) and the fitted parametrized pdfs from eq. (\ref{eq:fit}) (orange lines) for the base 10 logarithms of the chemical mixing ratios for Planet 17 in the database. The ground truth second quartiles (50-th percentiles) in this case were 
-6.46, -10.31,  -5.01,  -5.91 and -9.39, 
respectively. 
}      
\label{fig:fits}
\end{figure}

The posterior pdf (probability density function)  for each of the seven target parameters was suitably parameterized with an analytical ansatz. 
 The ansatz assumed no correlations for the chemical mixing ratios and represented each one as a weighted sum of a Gaussian distribution and a uniform distribution:

\begin{eqnarray}
f_i(x_i;\alpha_i,\mu_i,\sigma_i,m_i)&=&
\alpha_i\times \frac{1}{\sigma_i \sqrt{2 \pi}} e^{-\left(x_i-\mu_i\right)^2/2 \sigma_i^2}
\nonumber
\\
&+&\left(1-\alpha_i\right) \times\frac{\Theta\left(x_i+12\right) \Theta\left(m_i-x_i\right)}{m_i+12},
\label{eq:fit}
\end{eqnarray}
where $x_i=\log X_i$ is the base 10 logarithm of the mixing ratio for the $i$-th chemical absorber, and $\alpha_i$ and ($1-\alpha_i$) are the fractional contributions of the Gaussian and the uniform distribution, respectively, to the posterior distribution $f_i$ for the $i$-th absorber. The remaining quantities in eq.~(\ref{eq:fit}) are as follows: $\mu_i$ and $\sigma_i$ are respectively the mean and the standard deviation of the Gaussian, and  $m_i$ is the upper endpoint of the uniform distribution (the lower endpoint is taken as $-12$, hence the range is $m_i-(-12)=m_i+12$). This ansatz has a good physical justification, which is illustrated in Figure~\ref{fig:fits}, where we compare the given ground truth distributions (blue histograms) and the fitted parametrized pdfs from eq. (\ref{eq:fit}) (orange lines) for the log-mixing ratios $x_i$ for Planet 17 in the database. 
We observe that when the concentration of the corresponding absorber is relatively large ($x_i \gtrsim -6$), the absorber is well detected and the posterior distribution resembles a Gaussian centered near the true $x_i$ value, as shown in the first, third and fourth panel of Fig.~\ref{fig:fits}. If, on the other hand, the concentration is rather small (or alternatively, if the absorber is rather weak), the presence of this absorber is not detected and the posterior distribution is uniform up to some detection threshold $m_i$ (see the fifth panel in Fig.~\ref{fig:fits}). In intermediate cases like the one in the second panel, the posterior distribution can be approximated with a linear sum of those two shapes.
The parameters of the ansatz (\ref{eq:fit}) were fit to the training data by minimizing the squared difference between the cumulative distributions of the parametrized distribution and the provided ground truth distributions on 1000 points. The minimization was done using the {\sc scipy} minimize function with the L-BFGS-B method \cite{lbfgsb1,lbfgsb2,Scipy2020}. 

We noticed that in contrast to the chemical abundances, for the planet radius $R_p$ and the planet temperature $T$, the corresponding posterior distributions in the ground truth training data exhibited a relatively high degree of correlation. To reflect this fact, we used a different ansatz, namely, a bivariate normal distribution 
\begin{equation}
f_{R_pT}\left(R_p, T;\mu,\sigma_{R_p},\sigma_T,\rho\right)=\frac{\exp \left(-\frac{1}{2}\left(\left[R_p, T\right]-\mu\right)^T \Sigma^{-1}\left(\left[R_p, T\right]-\mu\right)\right)}{2 \pi|\Sigma|},
\label{eq:TPansatz}
\end{equation}
where $\mu = \left[\mu_{R_p} , \mu_T  \right]$ contains the respective means and
$$\Sigma=\left(\begin{array}{cc}
\sigma_{R_p}^2 & \sigma_{R_p} \sigma_T\rho \\
\sigma_{R_p} \sigma_{T }\rho & \sigma_T^2
\end{array}\right)
$$
is defined in terms of the respective standard deviations $\sigma_{R_p}$ and $\sigma_{T}$ and the correlation coefficient $\rho$. The correlation coefficient was taken as $\rho=-0.7$ and the remaining four parameters of the ansatz (\ref{eq:TPansatz}) were then fit to the ground truth training data by computing the covariance matrix with the {\sc NumPy} package \cite{numpy}.

\begin{figure}[t]
\centering
\includegraphics[width=0.19\textwidth]{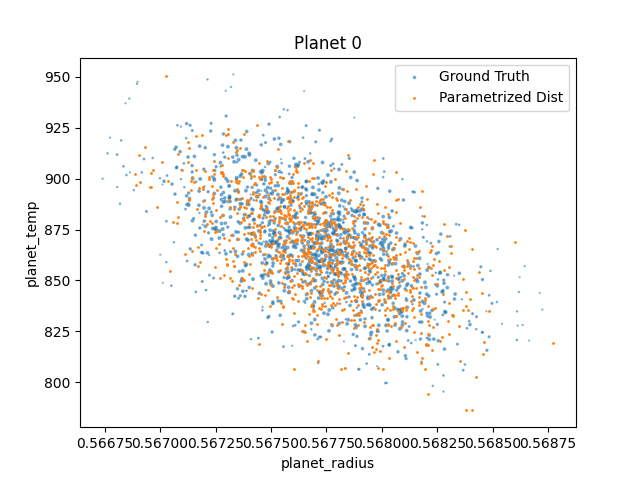}
\includegraphics[width=0.19\textwidth]{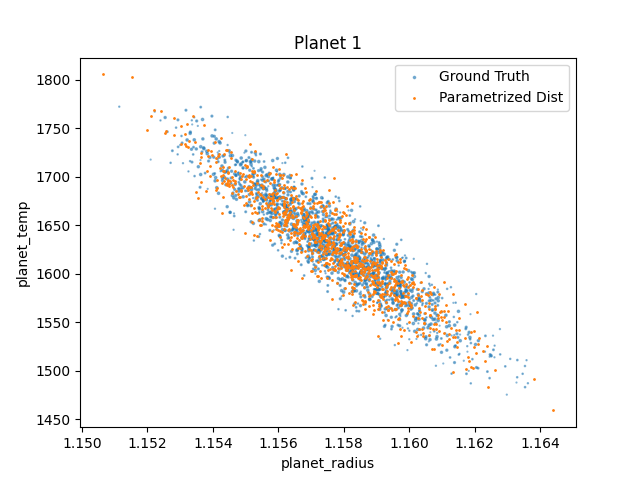}
\includegraphics[width=0.19\textwidth]{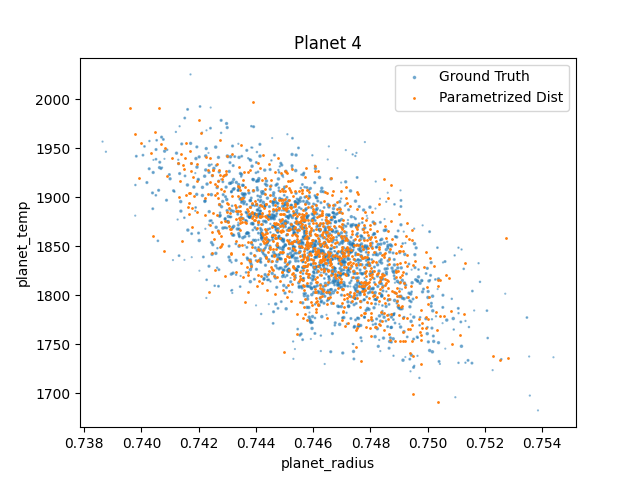}
\includegraphics[width=0.19\textwidth]{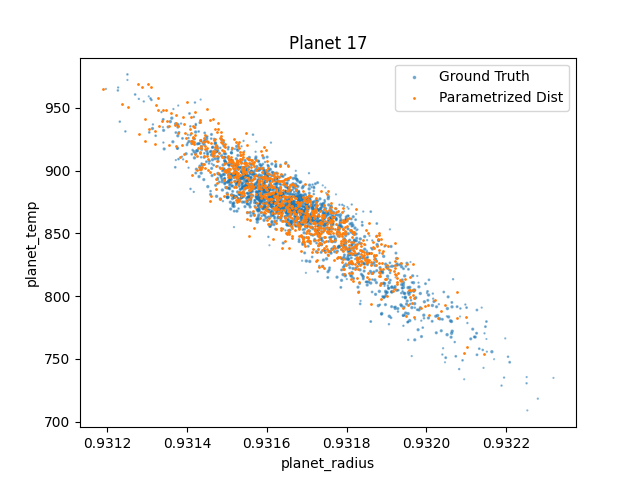}
\includegraphics[width=0.19\textwidth]{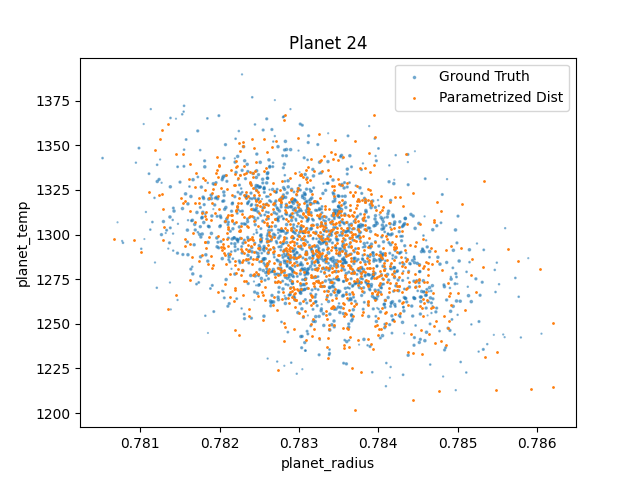}     
\caption{Probability distributions for the planet radius ($x$-axis) and the planet temperature ($y$-axis) for five different planets in the database. The blue points are sampled from the given ground truth distributions, while the orange points are sampled from the respective fitted parameterized distributions (\ref{eq:fit}). 
\label{fig:comparison}
}
\end{figure}

The parametrization of the posterior distributions was tested in two ways. First, a qualitative comparison of the two distributions was performed by visual inspection; a few representative examples are shown in Fig.~\ref{fig:comparison}, and they all exhibit good agreement. A second quantitative test involved computing the two-sample Kolmogorov–Smirnov (KS) test scores between the ground truth distributions and the respective parametrized ansatz. (The KS score is the largest absolute difference between the cumulative distribution functions of the two distributions under consideration.) The corresponding average scores are shown in Table~\ref{tab:KS} and are observed to be small, which indicates good agreement, since a lower KS score implies a better fit.

\begin{table}[htb]
\centering
\renewcommand\arraystretch{1.5}
\caption{Average two-sample KS test scores between the ground truth distributions and the parametrized distributions (\ref{eq:fit}) and (\ref{eq:TPansatz}).
}
\label{tab:KS}
\begin{tabular}{|l|l|l|l|l|l|l|l|}
\hline
Parameters & $R_p$   & T    & $H_2O$  & $CO_2$  & $CO$   & $CH_4$  & $NH_3$  \\ \hline
Test Score & .036 & .042 & .034 & .038 & .042 & .035 & .033 \\ \hline
\end{tabular}
\end{table}

\subsection{Model architecture}
\label{sec:architecture}

Three separate ML models were used to obtain final predictions. Two of those models were used to obtain the four trainable ansatz parameters in (\ref{eq:TPansatz}) for the planet radius $R_p$ and temperature $T$. These two models were composed of fully connected neural networks.  The final activation function was a sigmoid and its output was converted to the respective ansatz parameter by a linear transformation involving the range and minimum value for each parameter. 

\begin{figure}[t]
\includegraphics[width=0.99\textwidth]{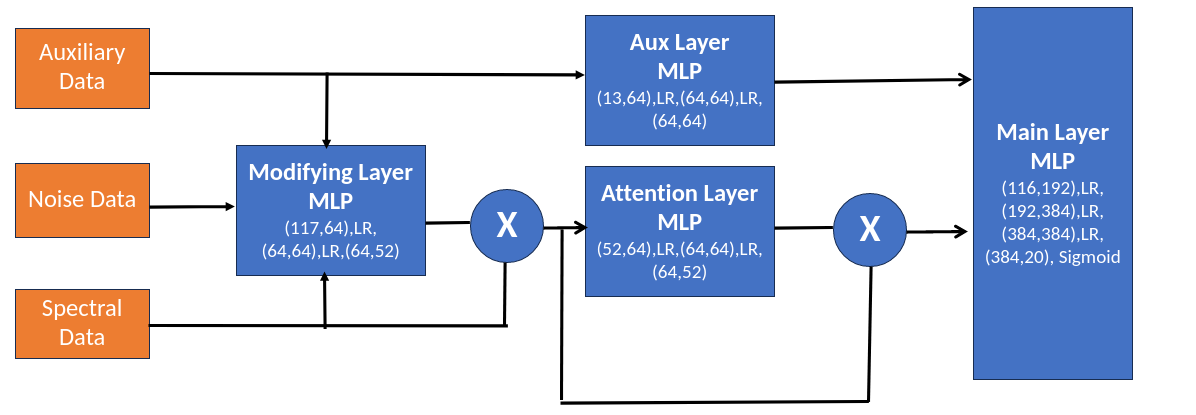}     
\caption{The model architecture used for the chemical model. LR represents the Leaky ReLU activation function. The $\times$ symbol indicates an elementwise multiplication of two data sets with the same size.}
\label{fig:architecture}
\end{figure}

For the ML model predicting the chemical abundances, we used a different architecture shown in Fig.~\ref{fig:architecture}. The model used several fully connected neural networks arranged into four modules represented with the blue boxes in the figure. As indicated with the $\times$ symbol, the outputs of the modifying layer modules were used to multiplicatively reweight data from a previous stage. 
The model's final activation function was a sigmoid. The model outputs 20 values which get reshaped into a $(4,5)$ matrix. In order to convert the sigmoid output into a meaningful value for the respective parameter, each column was multiplied elementwise  with the vector $[1,11,11,-6]$, then the vector $[0,-12,-12,6]$ was added to the product. This linear transformation keeps the prediction for $\alpha_i$ unchanged, but maps the predictions for $m_i$ and $\mu_i$ between $-12$ and $-1$ and the prediction for $\sigma_i$ between $0$ and $6$. These scaled values were then used as the predicted ansatz parameters, where each column contains the $[\alpha_i,m_i,\mu_i,\sigma_i]$ set of parameters for the log-mixing ratio of the $i$-th chemical absorber. 

\subsection{Loss function}

The evaluation metric for the data challenge involved the two-sample Kolmogorov-Smirnov test. Therefore, we chose our loss function to mimic the KS distance between the parametrized ground truth ansatz and our predicted distribution. For the planet radius and temperature, both of those distributions are Gaussian, and in that case the KS distance can be computed analytically in terms of the respective means and standard deviations.

Since the ansatz (\ref{eq:fit}) for the chemical abundances was more complicated, it was not possible to obtain an analytic expression for the KS distance. Instead, we used the following approximation
\begin{equation}
L(y,\hat y)= \sum_{i=0}^{5} 1.5 |\alpha_i-\hat \alpha_i| + \\|(1-\alpha_i)(m_i- \hat m_i)| + \alpha_i \sqrt { \left(\mu_i-\hat \mu_i \right)^2+ \left( \sigma_i - \hat \sigma_i \right)^2 },
\label{eq:ChemicalLoss}
\end{equation}
where the first (second) term enforces that the mean of the Gaussian (uniform) component of the posterior ansatz (\ref{eq:fit}) was predicted correctly, while the last term is proportional to the earth mover's distance between the respective Gaussian components only.\footnote{At the time of the competition, the code implementing the last term in (\ref{eq:ChemicalLoss}) contained a typo which has been fixed since.}

\subsection{Training}

\begin{figure}[t]
     \centering
     \includegraphics[width=0.9\textwidth]{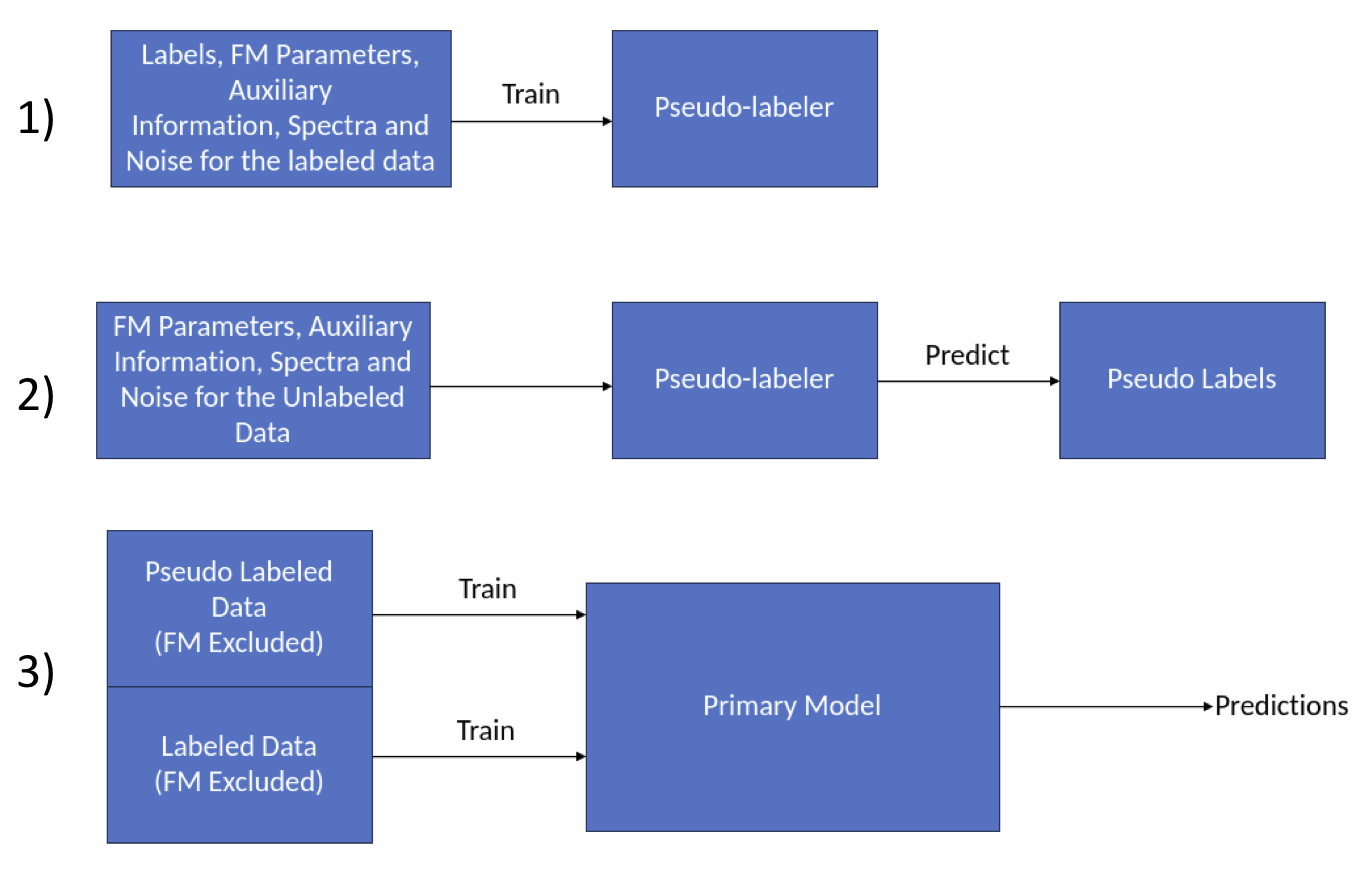}
     \caption{Schematic illustration of the semi-supervised approach used for the temperature ML model.}
     \label{fig:semisupervised}
\end{figure}

The models were constructed, trained and tested using the {\sc PyTorch} package \cite{torch}. The training was performed with the {\sc ADAM} optimizer \cite{ADAM}. For each model, we used different learning rates throughout the three different stages of training, as well as different number of epochs for each stage. The training of the $R_p$ model proceeded over 70 epochs, divided into stages of 20, 20 and 30, with learning rates $0.0001$, $0.00025$, and $0.0001$, respectively. The training of the temperature model was similar, with learning rates of $0.0005$, $0.0005$, and $0.0001$, respectively. Finally, the chemical composition model was trained over 50 epochs, split into stages of 20, 25 and 5, with the same learning rates as the temperature model.

\begin{figure}[t]
\centering
\includegraphics[width=.49\textwidth]{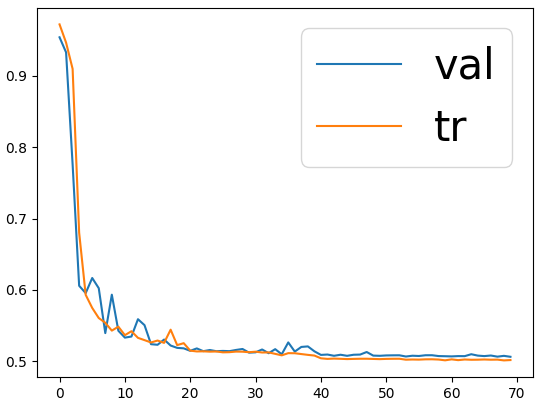}
\includegraphics[width=.49\textwidth]{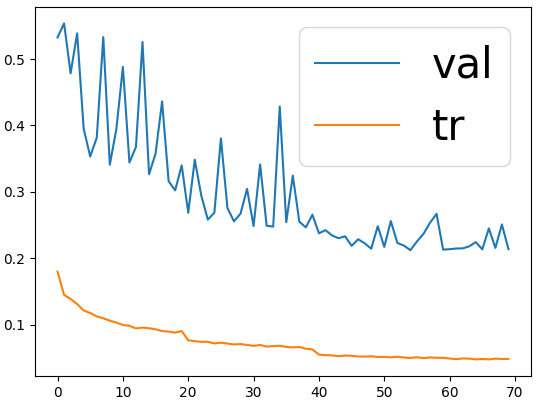}\\
\includegraphics[width=.49\textwidth]{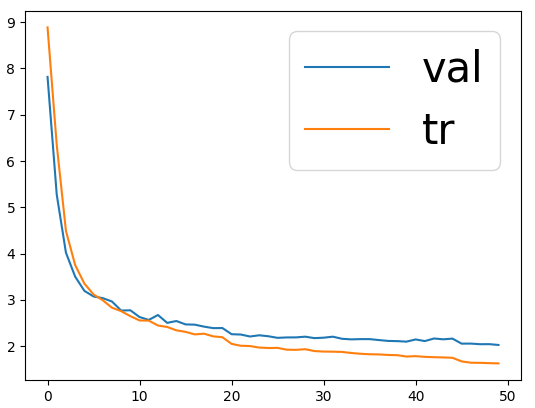}
\caption{Evolution of the training loss (orange lines) and the validation loss (blue lines) with the number of epochs on the horizontal axis, for the three ML models: for the planet radius $R_p$ (top left panel), for the atmospheric temperature $T$ (top right panel) and for the chemical abundances (bottom panel). Each model was trained on 80\% of the available training data and then validated on the remaining 20\%. 
} 
\label{fig:losses}
\end{figure}

In order to take advantage of the large amount of unlabelled training data available, we adopted the three-step semi-supervised approach illustrated in Fig.~\ref{fig:semisupervised}. In the first step, we develop a secondary ML model whose purpose is to provide pseudolabels for the unlabelled data. This model is trained on the available labelled data. More importantly, among its inputs, it additionally includes the provided true values for the forward model parameters ($R_p$, $T$, and $x_i$) used for the generation of the training spectra with the {\sc TauRex} forward model. In the second step, this trained ``pseudo-labeler'' is applied on the unlabelled data to predict labels for it. In the final third step, the primary ML model described in Section~\ref{sec:architecture} above is trained on a weighted mixture of labelled and pseudo-labelled data, where the contribution of the latter is suppressed by a factor $\beta$. Note that the primary ML model does not make any use of the forward model parameters, since they are not available for the competition test data.

For the training, we used a train-test split of 4:1 (80\% - 20\%). The evolution of the respective losses with the number of epochs for each of the three models is shown in Fig.~\ref{fig:losses}. 

\section{Results}
\label{sec:results}

\begin{table}[t]
\centering
\renewcommand\arraystretch{1.5}
\caption{Two-sample KS test scores on the validation set for four different model configurations.}
\label{tab:w_table}
\begin{tabular}{|c|c|c|c|c|l|l|l|l|l|l|l|l|}
\hline
\multicolumn{5}{|c|}{Model configuration} & \multicolumn{7}{c|}{Scores} \\
\hline
\ No\ &\ SSL\ &\ Train\ &\ Test\ & $~~\beta~~$ & $R_p$  & T      & $H_2O$ & $CO_2$ & $CO$   & $CH_4$ & $NH_3$ \\ \hline
I  & No  & 80\%  & 20\% & ---      & 0.5064 & 0.3129 & 0.1780 & 0.1558 & 0.1284 & 0.1807 & 0.1739 \\ \hline
II & Yes & 80\%  & 20\% & 5        & 0.5064 & 0.2142 & 0.1782 & 0.1558 & 0.1284 & 0.1806 & 0.1741 \\ \hline
III& Yes & 80\%  & 20\% & 1        & 0.5064 & 0.2096 & 0.1778 & 0.1558 & 0.1284 & 0.1809 & 0.1741 \\ \hline
IV & Yes & 100\% & 20\% & 5        & 0.5030 & 0.1606 & 0.1485 & 0.1129 & 0.1004 & 0.1410 & 0.1588 \\ \hline
\end{tabular}
\end{table}

\begin{figure}[t]
\centering
\includegraphics[width=0.9\textwidth]{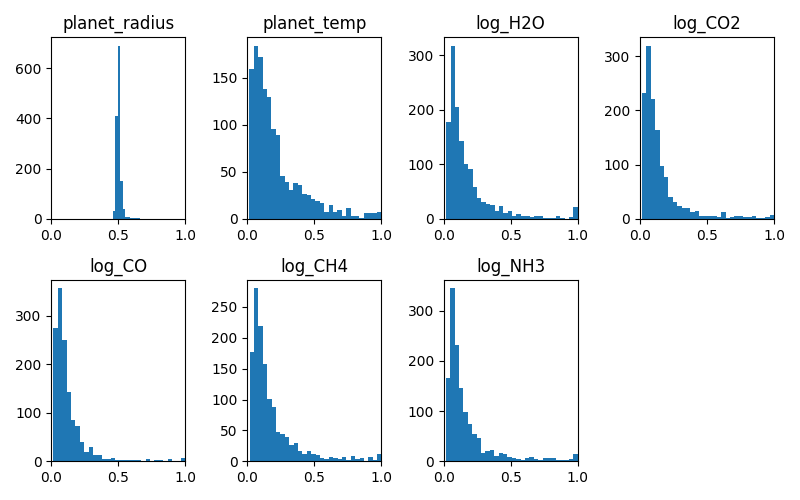}
\caption{Histograms of two-sample KS test scores of the planets in the validation set for each of the seven target parameters. }
\label{fig:w_dist}
\end{figure}

Different hyperparameter configurations were tested and the optimal configuration was selected based on the results on the validation set. Several representative model configurations are shown in Table~\ref{tab:w_table}, together with the corresponding average two-sample KS test scores on the validation set for each of the seven target parameters. Comparing the results for model configurations I and II in the table, we note that the semi-supervised learning (SSL) approach of Fig.~\ref{fig:semisupervised} made a notable improvement only in the case of the temperature ML model. Therefore, for simplicity, the chemical and the planet radius models used only labeled data for training. We checked that different reweightings of the pseudo-labelled training data (model configuration III is one such choice) did not significantly impact the results either. 

The final models were trained with the selected optimal hyperparameter configurations using all of the training data (model configuration IV in Table~\ref{tab:w_table}). The resulting distributions of two-sample KS test scores by parameter are shown in Fig.~\ref{fig:w_dist}.

\section{Analysis}

\begin{figure}[t]
     \centering
     \includegraphics[width=\textwidth]{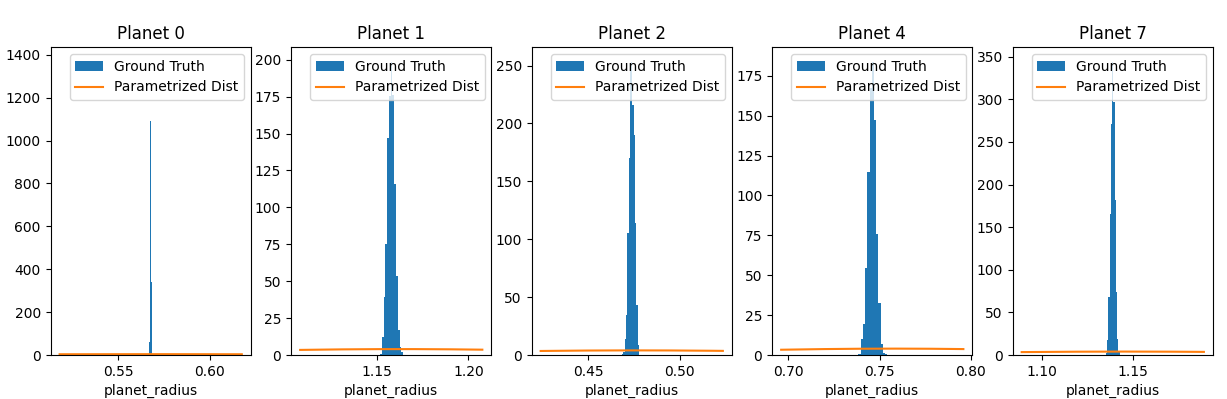}
     \caption{Comparison of the ground truth posterior distributions for $R_p$ (blue histograms) with the model's prediction (orange histograms) for several representative planets.} 
     \label{fig:R_pred_val}
\end{figure}

The predictions from the ML model configuration IV in Table~\ref{tab:w_table} were entered in the 2023 Ariel Machine Learning Data Challenge and ended up being among the winning solutions. In the process, some interesting lessons have been learned.

{\bf The difficulty in reproducing the posterior distributions for the planet radius $R_p$.}
The results shown in Table~\ref{tab:w_table} and Figure~\ref{fig:w_dist} reveal that while the posterior distributions for six of the seven target parameters were reproduced quite well, the model was struggling with the case of the planet radius $R_p$, where the KS distance did not drop below 0.5. The reason for this is illustrated in Figure~\ref{fig:R_pred_val}, which compares the ground truth posterior distributions for $R_p$ (blue histograms) with the model's prediction (orange histograms) for several representative planets in the data. One can observe that while the averages of the two distributions (ground truth and model prediction) are consistent, the standard deviations are very different, and the prediction (orange lines) is always much wider than the ground truth. This actually explains why the KS distance is of order 0.5: this is the asymptotic answer for two distributions with similar means and very different variances. The extreme narrowness of the ground truth distributions for $R_p$ poses a significant challenge for the ML model. For example, whenever the model's prediction for the mean is outside the ground truth distribution, the best the model can do is to inflate the standard deviation and thus ensure some overlap between the two distributions. This phenomenon is precisely what is seen in Figure~\ref{fig:R_pred_val}.

We also note that in many instances, the forward model value of $R_p$ used to generate the spectrum lies completely outside the ground truth posterior distribution, indicating that the Bayesian retrieval procedure was too overconfident with regards to the $R_p$ parameter. This motivates further investigations, e.g., applying ML for retrieving the forward model parameters directly via regression techniques.

{\bf The role of the sampling distribution used for creating the training dataset.} Recently, the impact of the choice of sampling distribution on the information content in the training database  has been under intense discussion \cite{2022ApJ...934...31F}. Interestingly, we note that the average scores for the chemical abundances seem to be correlated to the sampling range used for each chemical absorber. For example, the log-mixing ratios for $H_2O$, $NH_3$ and $CH_4$ share the same sampling range (from -9 to -3) and correspondingly, have similar mean scores ($\approx 0.18$). On the other hand, $CO_2$ was sampled from -9 to -4, and its mean score is better, on the order of 0.15. Lastly, $CO$ was sampled from -6 to -3, and has the best score, on the order of 0.13. While there may be other factors at play as well, these results suggest that the retrieval outcome may depend not only on the choice of function for the sampling distribution, but also on the sampling range of parameters when creating the training and testing datasets. 

{\bf The benefit from semi-supervised learning in enlarging the effective training dataset.} Semi-supervised techniques allow for leveraging all of the available training data (labelled and unlabelled). They are especially useful when labelling the data is prohibitively costly or simply not possible (for example, the competition rules did not allow using {\sc Taurex} to label the unlabelled data). In this year's Ariel data challenge, the majority of the available training data was unlabelled, which motivated the semi-supervised approach described in Figure~\ref{fig:semisupervised}. As discussed in Section~\ref{sec:results}, this noticeably improved the score for the planet temperature,  which was not handled as well by the regular supervised method. On the other hand, the supervised approach already worked quite well for the chemical abundances, and the addition of pseudo-labelled training data led to marginal improvements. Finally, in the case of the planet radius, the limiting factor was not the lack of training data, as already discussed above.

Comparing Table~\ref{tab:w_table} to Table~\ref{tab:KS}, we can see that our scores are not close to the optimal score which would occur if the ansatz parameters were predicted perfectly. Therefore, we believe that the approach described here, while successful in this year's competition, can be perfected further, e.g., by including more features motivated by domain knowledge, by considering a more general ansatz allowing correlations among the chemical abundances, improved model expressiveness and loss functions. We look forward to next year's Ariel Machine Learning Data Challenge.

\subsubsection{Acknowledgements} We thank the organizers of the Ariel 2023 Data Challenge for designing and conducting such an exciting and insightful competition and for continuous support and encouragement throughout. We are also grateful to the anonymous referees for valuable comments and constructive suggestions on the manuscript.

%
%
%
 \bibliographystyle{splncs04}
 \bibliography{refs}
%
%



\end{document}